\documentclass[aps,twocolumn,showpacs,floatfix]{revtex4-1}
\usepackage{outlines} 

\usepackage[caption=false]{subfig}
\usepackage{xcolor}
\usepackage{verbatim}
\usepackage{graphicx}
\usepackage{dcolumn}
\usepackage{bm}
\usepackage{color}
\usepackage[colorlinks=true, allcolors=blue]{hyperref}
\usepackage{makeidx}
\usepackage{amsmath}
\usepackage{amssymb}
\usepackage{mathrsfs}
\usepackage{soul} 

\makeindex
\def\>{\right\rangle}
\def\<{\left\langle}
\def\be{\begin{equation}}
\def\ee{\end{equation}}
\def\ba{\begin{aofprray}{l}}
\def\ea{\end{aofprray}}

\def\beq{\begin{eqnarray}}
\def\eeq{\end{eqnarray}}

\usepackage{soul}

\begin{document}
	\title{Quench-induced entanglement and relaxation dynamics in Luttinger liquids}
	\author{Alessio Calzona$^{1,2,3}$, Filippo Maria Gambetta$^{1,2}$, Fabio Cavaliere$^{1,2}$, Matteo Carrega$^{4}$, and Maura Sassetti$^{1,2}$}
	\affiliation{ $^1$ Dipartimento di Fisica, Universit\`a di Genova, Via Dodecaneso 33, 16146, Genova, Italy.\\
		$^2$ SPIN-CNR, Via Dodecaneso 33, 16146, Genova, Italy.\\
		$^3$ Physics and Materials Science Research Unit, University of Luxembourg, 162a avenue de la Fa\"iencerie, L-1511 Luxembourg. \\ 
		$^4$ NEST, Istituto Nanoscienze-CNR and Scuola Normale Superiore, Piazza San Silvestro 12, I-56127 Pisa, Italy.
	} 
	\date{\today}
	\begin{abstract}
		We investigate the time evolution towards the asymptotic steady state of a one dimensional interacting system after a quantum quench. We show that at finite times the latter induces entanglement between right- and left- moving density excitations, encoded in their cross-correlators, which vanishes in the long-time limit. This behavior results in a universal time-decay $ \propto t^{-2} $ of the system spectral properties, in addition to non-universal power-law contributions typical of Luttinger liquids. Importantly, we argue that the presence of quench-induced entanglement clearly emerges in transport properties, such as charge and energy currents injected in the system from a biased probe, and determines their long-time dynamics. In particular, the energy fractionalization phenomenon turns out to be a promising platform to observe the universal power-law decay $ \propto t^{-2} $ induced by entanglement and represents a novel way to study the corresponding relaxation mechanism.
	\end{abstract}
	\maketitle
\section{Introduction}
The study of non-equilibrium dynamics of many-body quantum systems is one of the most challenging and long standing problems in various fields of physics.
Here, very different fundamental aspects can be mentioned, including photo-induced biological processes~\cite{Engel:2007,Imamoglu:2015},
formation of strongly-correlated bound-states~\cite{Brachmann:1997,Berges:2004}, quantum phase transitions~\cite{Dziarmaga:2010,Polkovnikov:2011,Bastidas:2012},
relaxation and equilibration dynamics~\cite{Eisert:2015,DAlessio:2016,Essler:2016}. Among all, the recent technological developments in atomic and condensed
matter physics offer a very promising platform to investigate these important issues. Nowadays it is indeed possible to prepare a quantum system in a given
non-equilibrium state and to study its evolution in real time, for instance by using cold atoms~\cite{Kinoshita:2006,Bloch:multi,Cheneau:2012,Trotzky:2012,Langen:2015},
trapped ions~\cite{Blatt:2012,Jurcevic:2014,Richerme:2014}, or even quantum conductors in solid-state devices~\cite{Milletari:2013,Inoue:2014}.
An intriguing possibility to drive a quantum system out of equilibrium is to perform a \emph{quantum quench}, i.e. a sudden change in time of some of its parameters~\cite{Cazalilla:2006,Calabrese:2006,Cazalilla:2016}. Such a procedure is available in state-of-the-art systems of cold atoms~\cite{Bloch:multi},
in which transport and real-time control experiments have been recently reported~\cite{Cheneau:2012,Trotzky:2012,Langen:2015,Brantut:2012,Krinner:2015,Husmann:2015}.
The possibilities offered by the latter setups have given a significant boost to theoretical research, especially in the case of isolated one-dimensional (1D) integrable systems. Due to the presence of an infinite number of locally conserved quantities and the unitary time evolution, a fundamental question is if this kind of systems does relax to a steady state and, if so, how. In the seminal paper by Rigol et al.~and subsequent works~\cite{Rigol:GGE,Vidmar:2016} it was shown that this is the case if one focuses on local observables. Moreover, it was conjectured that the state reached by the system after the quench is locally described by a non-thermal density matrix. The latter can be obtained within the generalized Gibbs ensemble (GGE), which takes into account the presence of the local conserved quantities. Different experimental results and theoretical works have demonstrated the existence of this \emph{prethermalization} regime for various models~\cite{Berges:2004,Langen:2015gge}, although a general proof is still lacking. In the context of interacting 1D systems, it is well established that their low-energy equilibrium properties are described by the Luttinger liquid (LL) model~\cite{Voit:1995,vonDelft:1998,Giamarchi:2004}. They exhibit peculiar effects stemming from their non-Fermi liquid nature due to inter-particle interactions, such as charge and spin fractionalization~\cite{Giamarchi:2004,Jompol:2009,Deshpande:2010,Safi:1995,Auslaender:2005,Steinberg:2010,Deshpande:2010,Kamata:2014,Calzona:2015,Calzona:2016,Milletari:2013,Inoue:2014,Perfetto:2014,Karzig:2011}. In recent years the LL model has also been proven to be a very powerful tool for studying the dynamics of 1D systems after a quench of the interaction strength. In particular, the subsequent relaxation
towards a steady-state and the characterization of the latter has been the focus of many recent works ~\cite{Cazalilla:2006,Cazalilla:2016,Perfetto:2011,Kennes:2013,Kennes:2014,Schiro:2014-15,Porta:2016,Gambetta:2016,Calzona:2017}.
   
In this paper we will concentrate on the transient regime following an interaction quench in a LL. We demonstrate the presence of entanglement between right- and left-moving density excitations, encoded in their cross-correlators, as argued in similar contexts~\cite{Calabrese:2006, Cazalilla:2016,Calabrese:2016}. We will show that this quench-induced entanglement vanishes in the steady-state, inducing a \emph{universal} power-law decay in the cross-correlators, i.e. independent of any of the quench parameters.
In order to highlight the presence of entanglement and its subsequent relaxation, we study its effects on observable properties of a 1D fermionic system. Specifically, we focus on the time evolution of the non-equilibrium spectral function (NESF) and identify in its long-time behavior a universal contribution $ \propto t^{-2} $, precisely due to the entanglement dynamics via cross-correlators. However, the latter is in strong competition with non-integer power-law decays typical of LLs. Since the universal power law $ \propto t^{-2} $ originates directly from the evolution of entanglement, one would expect signatures of the latter in the long-time behavior of the system observable properties. On the other hand, exponents of LL-like non-integer power laws usually strongly depend on the specific quantity under examination~\cite{Voit:1995,vonDelft:1998,Giamarchi:2004}. We therefore explore the transient dynamics of transport properties, which are strictly related to the NESF. In particular, we consider the injection process from an external probe and the subsequent dynamics of the LL after the quench, studying the injected charge and energy currents as a function of time. We demonstrate that for these quantities the universal character clearly emerges as the dominant contribution to the long-time behavior. Indeed, while the universal term $ \propto t^{-2} $ is present in both NESF and currents, non-integer LL-like power-law contributions feature greater exponents with respect to the NESF, leading to a clearer emergence of the universal character in transport properties.  Finally, we find that the latter is even more evident in the energy fractionalization ratio, which thus represents a very promising tool to probe the relaxation effects of entanglement.
	
The paper is organized as follows. In Sec.~\ref{sec:model} we analyze the time evolution of two-point correlators after an interaction quench in a generic 1D system with short-range inter-particle interactions. In Sec.~\ref{sec:spectral} we specialize to the case of the dynamics of spectral properties of a fermionic spinless LL after a quantum quench. In Sec.~\ref{sec:trasport} we consider the behavior of the charge and energy currents injected from an external probe into the LL. Finally, Sec.~\ref{sec:conclusions} summarizes our conclusions.

\section{Model and quench-induced entanglement}\label{sec:model}
Let us consider an interacting 1D system with short-range interactions. Assume that this is initially prepared in the ground state $ |\bm{0}_i\rangle $ of the initial
Hamiltonian $ H_i $. At $ t=0 $, the system is brought out-of-equilibrium by suddenly changing the strength of inter-particle
interactions. The subsequent dynamics is thus governed by the final Hamiltonian $ H_f $~\cite{Cazalilla:2006, Cazalilla:2016}. In particular, the two
Hamiltonians involved in the quench can be written as (hereafter $ \hbar=1 $)~\cite{Voit:1995,vonDelft:1998,Giamarchi:2004} 
\begin{equation}
	H_\mu=\frac{u_\mu}{2}\sum_{\eta=\pm}\int_{-\infty}^{\infty}\left[\partial_{x}\phi_{\mu,\eta}(x)\right]^2\,dx,\label{eq:H}
\end{equation}
with $\mu=i,f$. Here, $ u_\mu=v/K_{\mu} $ are the mode velocities, $ v $ is the bare velocity and $ K_\mu=[1+g_{\mu}/(\pi v)]^{-1/2} $ is the dimensionless
Luttinger parameter describing the strength of inter-particle interactions~\cite{footnote:g2g4,footnote:direct} (with $ K_\mu=1 $ representing the non-interacting
case).
The bosonic fields $ \phi_{\mu,\eta}(x) $ diagonalize $ H_\mu $ and are chiral and evolve, with respect to $H_\mu$, as $ \phi_{\mu,\eta}(x)=\phi_{\mu,\eta}(x-\eta
u_\mu t) $. They thus describe right-- ($ \eta=+ $) and left-- ($ \eta=- $) moving excitations. Chiral fields before and after the quench are related to
each other by the canonical transformation~\cite{Gambetta:2016,Calzona:2017} 
\begin{equation}
\phi_{f,\eta}(x)=\sum_{\ell=\pm}\theta_{(\ell\eta)}\phi_{i,\ell}(x),\label{eq:canonical}
\end{equation}
with
$ 2\theta_\pm=\sqrt{K_i/K_f}\pm\sqrt{K_f/K_i} $.
To investigate the dynamics induced by the quench, it is useful to express the initial state $ |\bm{0}_i\rangle $ in terms of the \emph{final} chiral fields
$\phi_{f,\eta}(x)$. One obtains~\cite{Silva:2008,Sotiriadis:2014}
\begin{equation}
	|\bm{0}_i\rangle\propto\exp\left\{\sigma \int_{-\infty}^{\infty}[\partial_x \phi_{f,+}(x)]\phi_{f,-}(x)\,dx\right\}|\bm{0}_f\rangle,\label{eq:GSi}  
\end{equation}
with $ \sigma=\theta_{-}/\theta_{+}$ and $ |\bm{0}_f\rangle $ the ground state of $ H_f $. 
This implies that bosonic chiral fields $ \phi_{f,+}(x) $ and $ \phi_{f,-}(x)
$ are strongly entangled. It is thus interesting to inspect how the evolution of this entanglement affects the dynamics of the system itself. To this end, since $ |\bm{0}_i\rangle $ is a gaussian state, we can focus on the two-point correlators
\begin{align}
	D_{\alpha,\beta}(\xi;t,\tau)&\equiv2\langle \phi_{f,\alpha}(x-\xi,t-\tau) \phi_{f,\beta}(x,t)\rangle_i\nonumber\\
	&-\langle\phi_{f,\alpha}(x-\xi,t-\tau)\phi_{f,\beta}(x-\xi,t-\tau)\rangle_i\nonumber\\
	&-\langle\phi_{f,\alpha}(x,t)\phi_{f,\beta}(x,t)\rangle_i ,
	\label{eq:2points}
\end{align}
which fully characterize the system. Here, $ \langle\dots\rangle_i $ denotes the quantum average on $ |\bm{0}_i\rangle $, the initial ground state before the quench. Note that space-translational invariance is preserved by an interaction quench, i.e. the two-point correlators only depend on the relative coordinate $ \xi $. Importantly, the time-translational invariance is broken by the quench. In the absence of the latter - i.e. $ \sigma=0 $ in Eq.~\eqref{eq:GSi} - quantum averages in Eq.~\eqref{eq:2points} would be evaluated on the ground state $ |\bm{0}_f\rangle $, resulting in $ D_{\alpha,\beta}(\xi;t,\tau)\equiv
D^{\mathrm{eq}}_{\alpha,\alpha}(\xi;\tau)\delta_{\alpha,\beta} $. In this case, due to time-translational invariance, two-point correlators would only depend on the relative temporal coordinate $ \tau $ and fields with opposite chirality would not be entangled, as expected in equilibrium.\\
To investigate the dynamics induced by entanglement, we focus on the long-time behavior of two-point correlators $ D_{\alpha,\beta}(\xi;t,\tau) $, defined by $ t\gg\tau,\,\xi/u_f $. In this regime the evolution of the bosonic fields is governed by the post-quench Hamiltonian $ H_f $ and thus $ \phi_{f,\eta}(x,t)=\phi_{f,\eta}(x-\eta u_f t) $. With straightforward calculations the two-point correlators of Eq.~\eqref{eq:2points} evaluate to
\begin{widetext}
\begin{equation}
D_{\alpha,\beta}(\xi;t,\tau)=\sum_{\eta=\pm}\frac{\theta_{(\eta\alpha)}\theta_{(\eta\beta)}}{2\pi}\text{ln}\left\{\frac{[a-i \eta (\beta-\alpha)u_f t][a-i \eta (\beta-\alpha)u_f (t-\tau)]}{[a-i \eta (-\xi+(\beta-\alpha)u_f t + \alpha u_f \tau)]^2}\right\}. \label{eq:Daq}
\end{equation}
\end{widetext}
Here, the averages are computed by exploiting the canonical transformation between initial and final chiral fields of Eq.~\eqref{eq:canonical} and the relation~\cite{Voit:1995,vonDelft:1998,Giamarchi:2004} 
\begin{equation}
\label{eq:standardcorrelation}
\langle\phi_{i,\alpha}(x)\phi_{i,\beta}(y)\rangle_i=\frac{\delta_{\alpha,\beta}}{2\pi}\ln\left[\frac{L}{2\pi}\frac{1}{a-i\alpha(x-y)}\right],
\end{equation}
with $ L $ the length of the system and $ a $ a short-length cutoff. In particular, for $ \alpha=\beta $ one gets $ D_{\alpha,\alpha}(\xi;t,\tau)\equiv D_{\alpha,\alpha}(\xi;\tau) $,
i.e., breaking of time-translational
invariance does not affect auto-correlators, although they are different from their equilibrium counterparts. On the other hand, cross-correlators exhibit an explicit time dependence, 
\begin{multline}
D_{\alpha,-\alpha}(\xi;t,\tau)=\frac{\theta_{+}\theta_{-}}{2\pi}\\
\times\text{ln}\left\{\frac{[a^2+ 4 u_f^2 t^2][a^2+4 u_f^2 (t-\tau)^2]}{[a^2+ (-\xi-\alpha u_f (2t-\tau))^2]^2}\right\},
\label{eq:crosscorrelators}
\end{multline}
encoding the entanglement, and its decay in time, between bosonic fields $ \phi_{f,+}(x) $ and $ \phi_{f,-}(x)$. Note that cross-correlators are different from zero at any finite time $ t $ while $ D_{\alpha,-\alpha}(\xi;t,\tau)\rightarrow 0 $ for $ t\rightarrow \infty $. By expanding Eq.~\eqref{eq:crosscorrelators} in Taylor series in the long-time limit $ t\gg \tau,\, \xi/u_f $ one obtains a decay with integer power laws only, whose exponents are thus independent of the quench parameters. In particular, in the local case $ \xi=0 $ on which we will focus in the following one has 
\begin{equation}
D_{\alpha,-\alpha}(0;t,\tau)=\sum_{n=2}^{\infty}\frac{d_n(\tau)}{t^n}, \label{eq:Dabexpansionlocal}
\end{equation}
with coefficients $ d_n(\tau) $ independent of the chirality. Therefore, in the long-time limit, cross-correlators decay with a leading power-law behavior $ \propto t^{-2} $. Finite cross-correlators $ D_{\alpha,-\alpha}(\xi;t,\tau) $ are a hallmark of the quench-induced entanglement between the two counter-propagating bosonic fields and will determine the long-time relaxation of the system towards its steady-state. Moreover, due to their algebraic long-ranged behavior, one would expect observable signature of their decay in the system properties. 

\section{Time-dependent spectral function}\label{sec:spectral}
We now discuss the influence of quench-induced cross-correlations on a specific example of fermionic spinless LL with repulsive interactions ($K_\mu \leq
1$) and $K_i>K_f$. Using bosonization~\cite{Voit:1995,vonDelft:1998,Giamarchi:2004}, the system is described in terms of bosonic fields by Eq.~\eqref{eq:H} and the fermionic operator decomposes into right ($ R $) and left ($ L $) channels as $ \psi(x)=e^{i q_F x}\psi_R(x)+e^{-i q_F x} \psi_L(x) $. Here, $q_F$ is the Fermi wave-vector and 
\begin{equation}
\psi_r(x)=\frac{1}{\sqrt{2\pi a}} \exp[-i\sqrt{2\pi}\Phi_r(x)] , \label{eq:bosonization}
\end{equation} 
with $ r=R,L $. The bosonic field $ \Phi_r(x) $ can be expressed in terms of \emph{final} chiral fields as
\begin{equation}
\Phi_r(x)=\sum_{\eta}A_{(\epsilon_r\eta)}\phi_{f,\eta}(x),
\end{equation}
with $ 2A_{(\epsilon_r\eta)}=K_f^{-1/2}+\epsilon_r\eta K_f^{1/2} $ and $ \epsilon_{R/L}=\pm1 $.
To show how the decay of entanglement between opposite chiral excitations affects the dynamics of the system, we focus on the behavior of the local lesser Green function, 
\begin{equation}
G^{<}(t,t-\tau)\equiv i \langle\psi^\dagger(x,t-\tau)\psi(x,t)\rangle_i,
\end{equation}
in the regime $t>\tau$. Since the particle number is conserved, we can write
\begin{equation}
G^<(t,t-\tau)=G_R^{<}(t,t-\tau)+G_L^{<}(t,t-\tau),\label{eq:Glesserdef}
\end{equation}
where $ G_r^{<}(t,t-\tau) $ denotes the $ r- $channel lesser Green function. Using the bosonization identity of Eq.~\eqref{eq:bosonization} and recalling Eqns.~\eqref{eq:2points} and~\eqref{eq:Daq}, we obtain
\begin{equation}
G^{<}_r(t,t-\tau)=G_{r,\infty}^{<}(\tau)\mathcal{U}(t,\tau).
\label{eq:Glesseraq}
\end{equation}
Here,
\begin{align}
G_{r,\infty}^{<}(\tau)&=\frac{i}{2\pi a} e^{\pi \left[A^2_{\epsilon_r}D_{+,+}(0;\tau)+A^2_{-\epsilon_r}D_{-,-}(0;\tau)\right]}\nonumber\\
&=\left[\frac{a}{a + i u_f \tau}\right]^{\nu_+} \left[\frac{a}{a - i u_f \tau} \right]^{\nu_-}
\end{align}
represents the steady-state $ r- $channel lesser Green function while
\begin{align}
\mathcal{U}(t,\tau)&= e^{\pi A_+A_-[D_{+,-}(0;t,\tau)+D_{-,+}(0;t,\tau)]}\nonumber\\
&=\left\{\frac{[a^2+u_f^2(2t-\tau)^2]^2}{(a^2+4u_f^2t^2)[a^2+4u_f^2(t-\tau)^2]}\right\}^{\gamma}
\label{eq:U}
\end{align}
features the explicit time dependence encoded in the cross-correlators $ D_{\alpha,-\alpha}(0;t,\tau) $. Note that $ \mathcal{U}(t,\tau) $ does not depend on the channel index $ r $ and $ \mathcal{U}(t,\tau)\rightarrow1 $ for $ t\rightarrow\infty $. Here, $ \nu_\pm=\theta^2_\mp(A^2_++A^2_-) $ and $ \gamma=-A_+A_-\theta_+\theta_- $. In particular, one has $\gamma>0$ for the quench protocols with $K_i>K_f$ we are considering.  Importantly, the presence of cross-correlators
$D_{\alpha,-\alpha}(0;t,\tau)$ in the function $ \mathcal{U}(t,\tau) $ leads to a universal power-law decay of $ G_r^{<}(t,t-\tau) $ in the long-time limit. Indeed, by expanding $ \mathcal{U}(t,\tau) $ in Taylor series for $ \tau/t\ll 1 $ we obtain 
\begin{equation}
G^{<}_r(t,t-\tau)= G_{r,\infty}^{<}(\tau)\left[1+\sum_{n=2}^{\infty}\frac{g_n(\tau)}{t^n}\right].
\label{eq:Glesserr}
\end{equation}
Since in the local case we address here $ G^{<}_r(t,t-\tau) $ does not explicitly depend on the index $ r $, one readily obtains the long-time limit expansion of the full lesser Green function
\begin{align}
\label{eq:Glesser}
G^{<}(t,t-\tau)&=G^{<}_\infty(\tau)\left[1+\sum_{n=2}^{\infty}\frac{g_n(\tau)}{t^n}\right]\nonumber\\
&\approx G^{<}_\infty(\tau)\left(1+\frac{\gamma \tau^{2}}{2t^2}\right),
\end{align}
with $ G_\infty^{<}(\tau)=2G^{<}_{r,\infty}(\tau) $. Therefore, in the long-time limit, $ G^{<}(t,t-\tau) $ approaches its asymptotic value $ G^{<}_\infty(\tau) $ with a power-law decay $ \propto t^{-2} $, directly induced by the relaxation of cross-correlators $ D_{\alpha,-\alpha}(\xi;t,\tau) $ found in Eq.~\eqref{eq:Dabexpansionlocal}~\cite{footnote:nonlocal}.

The long-time behavior of Eq.~\eqref{eq:Glesser} immediately reflects on spectral properties, as one can see by inspecting the long-time limit of the local (lesser) NESF~\cite{Meden:1992,Kennes:2014,Nghiem:2017}
\begin{equation}
A^{<}(\omega,t)\equiv \frac{1}{2\pi}\int_{-\infty}^{\infty}e^{i\omega \tau}(-i)G^{<}(t,t-\tau)\,d\tau.
\label{eq:spectral}
\end{equation}
Indeed, as shown in Appendix~\ref{app:appendix}, we find
\begin{equation}
A^{<}(\omega,t)= \bar{A}_0\bigg[ \bar{A}_\infty^{<}(\omega) + \sum_{n=2}^{\infty}\frac{\mathcal{A}_n(\omega)}{t^n}+\frac{\mathcal{M}^A(\omega,t)}{t^\nu}\bigg],\label{eq:Alesser}
\end{equation}
with $ \bar{A}_0=(2\pi^2 v)^{-1} $ and all terms inside the square brackets dimensionless. Here, $ A_\infty^{<}(\omega)=\bar{A}_0\bar{A}_\infty^{<}(\omega) $ is the steady-state value of the NESF, already discussed in Refs.~\cite{Kennes:2014,Calzona:2017}. In this work, we focus on the time-decay of $ A^{<}(\omega,t) $ towards this asymptotic value. In particular, two distinct contributions emerge. The first one contains only integer power laws $ \propto t^{-n} $ (with $ n\geq2 $) and is entirely due to the decay of $ G^{<}(t,t-\tau) $ found in Eq.~\eqref{eq:Glesser}. Here, the coefficients present in the sum are given by $ \mathcal{A}_n(\omega)=2\pi^2 v\int_{-\infty}^{\infty}G^{<}_\infty(\tau) g_n(\tau)\, d\tau $, with $ g_n(\tau) $ defined in Eq.~\eqref{eq:Glesser}. We therefore obtain that the leading contribution of this term is a \emph{universal} power-law decay $ \propto t^{-2} $, regardless of, e.g., quench parameters. On the other hand, the second contribution contains the function $ \mathcal{M}^A(\omega,t) $ which, to the leading order in $ 1/t $, is an oscillating function with constant amplitude (see Appendix~\ref{app:appendix} for details). Thus, in the long-time limit, it decays with a LL-like \emph{non-universal} power law $ \propto t^{-\nu} $, with
\begin{equation}
\nu=\frac{K_f^4+K_i^2+3K_f^2(1+K_i^2)}{8K_f^2K_i}\geq1 \label{eq:nu}
\end{equation}
strongly dependent on quench parameters $ K_i $ and $ K_f $.
It turns out that the universal power-law behavior, which directly derives from the decay of entanglement between bosonic excitations $ \phi_{f,+}(x) $ and $ \phi_{f,-}(x)
$, is hardly visible in the transient of the NESF.
Indeed, for any reasonable quench one finds $ 1\leq\nu<2 $. Thus, the long-time decay of $ A^{<}(\omega,t) $ is governed by the non-universal contribution $ \propto t^{-\nu} $, with the universal one being a sub-leading term.
\begin{figure}
	\centering
	\includegraphics[width=\linewidth]{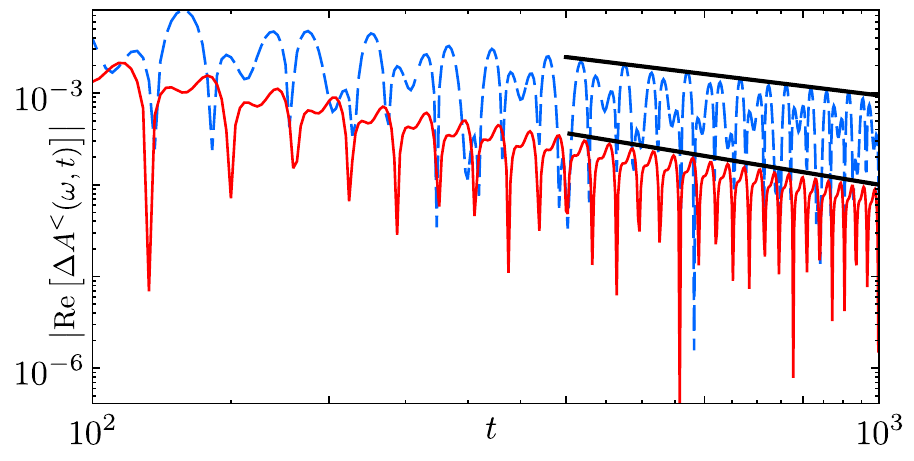}
	\caption{(Color online) Plot of $ |\text{Re}[\Delta A^{<}(\omega,t)]| $ [units $ \bar{A}_0=(2\pi^2 v)^{-1}$] as a function of time $ t $ [units $ (v q_F)^{-1} $] with $ \omega=-0.1\, vq_F $ for the quenches $ K_i=0.9\rightarrow K_f=0.7$ (blue, dashed) and $ K_i=0.8 \rightarrow K_f=0.4$ (red, solid).  Here, solid black lines represent the power-law behavior $ \propto t^{-\nu} $ for the two different cases.}
\label{fig:spectral}
\end{figure}
This is illustrated in Fig.~\ref{fig:spectral}, which shows the deviation of the lesser NESF from its steady-state value, $ \Delta A^{<}(\omega,t)=A^{<}(\omega,t)-A^{<}_\infty(\omega)
$, at large times and for two different interaction quenches~\cite{footnote:cutoff}. Here, the oscillating behavior due to $ \mathcal{M}^A(\omega,t) $ decays with non-universal power law $ \propto t^{-\nu} $ (see solid black lines) while no evidence of the universal behavior $ \propto t^{-2} $ is present.
Despite the sub-leading character of the universal contribution to the behavior of the NESF in Eq.~\eqref{eq:Alesser}, in the next Section we will demonstrate that it controls the long-time behavior of charge and energy currents in a transport setup. 

\section{Transient dynamics of transport properties}\label{sec:trasport}
\begin{figure}
	\centering
	\includegraphics[width=1\linewidth]{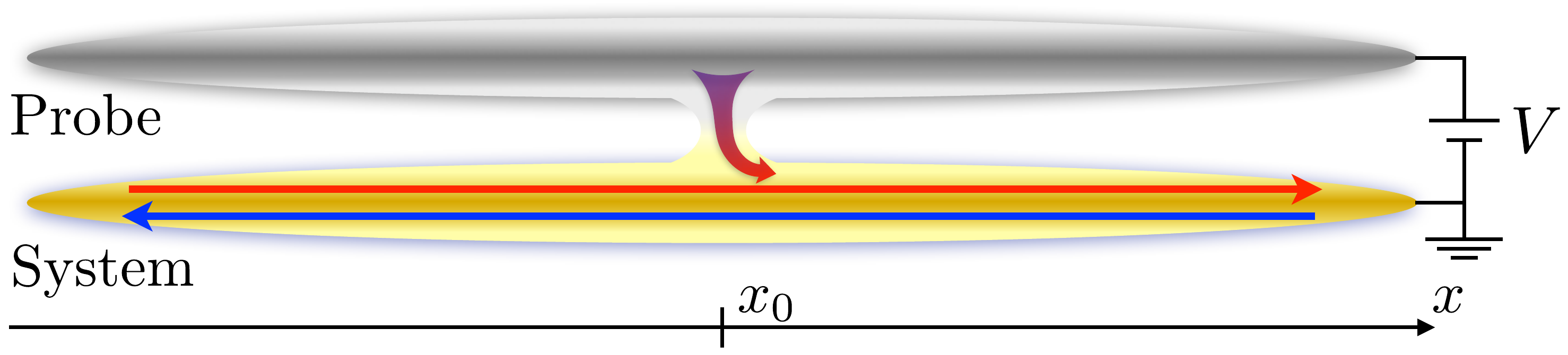}
	\caption{(Color online) Scheme of the system, modeled as a pair of counter-propagating channels, and the probe, biased with a dc voltage $ V $. At $ x=x_0 $, the probe injects $ R- $moving particles only.}
	\label{fig:setup}
\end{figure}

Assume now that immediately after the quench, the LL (hereafter dubbed \emph {the system}) is locally tunnel-coupled to a non-interacting 1D \emph{probe}, as sketched in Fig.~\ref{fig:setup}, described by the Hamiltonian
\begin{equation}
H_p=-iv\int_{-\infty}^{\infty}\chi^\dagger(x)\partial_x\chi(x)\,dx,
\end{equation} 
with $\chi(x)$ its fermionic field. The probe is subject to a bias voltage $ V $ measured with respect to the Fermi level of the system. We assume a local tunneling at $x_0$ which breaks inversion parity, focusing, e.g., on the injection in the system $ R- $channel only~\cite{Calzona:2017,Chevallier:2010,Dolcetto:2012,Vannucci:2015,Calzona:2017},
\begin{equation}
H_t(t)=\vartheta(t)\lambda \; \psi_R^\dagger(x_0)\chi(x_0)+\mathrm{h.c.}, \label{eq:Ht}
\end{equation}
where $ \lambda $ is the tunneling amplitude and $ \vartheta(t) $ is the Heaviside step function. The whole setup is assumed to be in thermal equilibrium before the quench, with $ \rho(0) $ the associated zero-temperature density matrix. 
We concentrate on chiral charge and energy currents, defined as 
\begin{align}
I_\eta(V,t) &= e\partial_t \int_{-\infty}^{\infty} \langle\delta n_\eta(x,t)\rangle \,dx,\label{eq:chiralI}\\
P_\eta(V,t) &= \partial_t \int_{-\infty}^{\infty} \langle\delta \mathcal{H}_\eta(x,t)\rangle \,dx. \label{eq:chiralP}
\end{align}
Here, 
\begin{align}
 n_\eta(x,t) &= -\eta\sqrt{\frac{K_f}{2\pi}}\partial_x\phi_{f,\eta}(x-\eta u_f t),\label{eq:chiraldensity}\\
 \mathcal{H}_\eta(x,t)&=\frac{u_f}{2}[\partial_x\phi_{f,\eta}(x-\eta u_f t)]^2\label{eq:chiralhamiltonian}
\end{align}
are the chiral particle and Hamiltonian densities, respectively, while  
\begin{equation}
\langle\delta \mathcal{O}(x,t)\rangle = \text{Tr}\{ \mathcal{O}(x,t)[\rho(t)-\rho(0)] \}
\end{equation}
represents the average variation, induced by the tunneling, of a given operator $ \mathcal{O}(x,t) $. The time-dependent full density matrix $ \rho(t) $ is evaluated in the interaction picture with respect to $ H_t(t) $~\cite{Calzona:2016,Calzona:2017}. The explicit expressions for the chiral charge and energy currents are computed, to the lowest order in tunneling amplitude and in the long-time limit, in Appendix~\ref{app:appendix}. These two quantities are directly related to the NESF in Eq.\ \eqref{eq:Alesser} and, not surprisingly, share with the latter an analogous structure 
\begin{align} I_\eta(V,t)&=\bar{I}_0\bigg[\bar{I}_\eta^\infty(V)+\sum_{n=2}^{\infty}\frac{\mathcal{I}_{\eta,n}(V)}{t^n}+\frac{\mathcal{M}_\eta^I(V,t)}{t^{\nu+1}}\bigg],\label{eq:current}\\ P_\eta(V,t)&=\!\bar{P}_0\bigg[\bar{P}_\eta^\infty(V)\!+\!\sum_{n=2}^{\infty}\frac{\mathcal{P}_{\eta,n}(V)}{t^n}\!+\!\frac{\mathcal{M}^P_\eta(V,t)}{t^{\nu+2}}\bigg],\label{eq:power}  
\end{align}
with $ \bar{I}_0=e|\lambda|^2 q_F(2\pi^2 v)^{-1}$, $ \bar{P}_0=|\lambda|^2 q_F^2(\pi^2 K_f )^{-1} $  and all the terms in square brackets dimensionless~\cite{footnote:cutoff}. In particular, Eqns.~\eqref{eq:current} and~\eqref{eq:power} consist of three contributions: a steady-state value~\cite{Gambetta:2016,Calzona:2017}, $ I^{\infty}_\eta(V)=\bar{I}_0\bar{I}^{\infty}_\eta(V) $ and $ P^{\infty}_\eta(V)=\bar{P}_0\bar{P}^{\infty}_\eta(V) $, respectively; one transient contribution that contains only integer power laws of time, stemming from the time dependence of $ G^{<}(t,t-\tau) $ in Eq.~\eqref{eq:Glesser}; another transient contribution which is related to the quenched LL \emph{non-universal} behavior. As well as $\mathcal{M}^A(\omega,t)$ in Eq.~\eqref{eq:Alesser}, both $\mathcal{M}_\eta^{I}(V,t)$ and $ \mathcal{M}^{P}_{\eta}(V,t)$ are functions whose leading term is an oscillating factor with constant amplitude. However, in sharp contrast with the NESF, here the non-universal transient contributions decay as $t^{-\nu-1}$ and $t^{-\nu-2}$ for the charge and energy currents respectively, with $\nu\geq1$ given in Eq.\ \eqref{eq:nu}. This behavior of the non-universal contribution is not surprising since, in essence, $A^<(V,t)\propto \partial_V \sum_{\eta}I_\eta(V,t) $ and $A^<(V,t)\propto \partial^2_V \sum_{\eta}P_\eta(V,t)$. Furthermore, since both quantities contain oscillating factors $\sim e^{iV t}$ in the functions ${\mathcal M}_\eta^{I,P}(V,t)$ [see Eqns.~\eqref{app:eq:MI} and~\eqref{app:eq:MP} in Appendix~\ref{app:appendix}], the exponents of the non-universal power laws will be modified according to the derivative over $V$. On the other hand, coefficients of the universal contributions $\mathcal{I}_{\eta,n}(V)$ and $\mathcal{P}_{\eta,n}(V)$ are independent of $ t $, and derivation with respect to $ V $ does not affect the universal power-law decay in time.
The relaxation dynamics of $I_\eta(V,t) $ and $P_\eta(V,t) $ are thus governed by the universal decay $ \propto t^{-2} $, which does not depend on the quench parameters and traces back to the behavior of cross-correlators $ D_{\alpha,-\alpha}(0;t,\tau) $ and thus of entanglement. The long-time dynamics of chiral charge and energy currents can thus directly reveal the entanglement between chiral bosonic fields $ \phi_{f,+}(x) $ and $ \phi_{f,-}(x) $ and its relaxation.

In particular, the chiral energy current represents a promising tool to elucidate the universal dynamics of the system. In fact, its non-universal transient contribution decays faster, leading to an earlier emergence of the $t^{-2}$ relaxation with respect to the charge current. In addition, the excess chiral energy current $\Delta P_\eta(V,t)=P_\eta(V,t)-P_\eta^\infty(V)$ displays a striking dependence on the chirality $\eta$, which turns out to help even more in detecting the universal behavior $\propto t^{-2}$. 
\begin{figure}
	\centering
	\includegraphics[width=\linewidth]{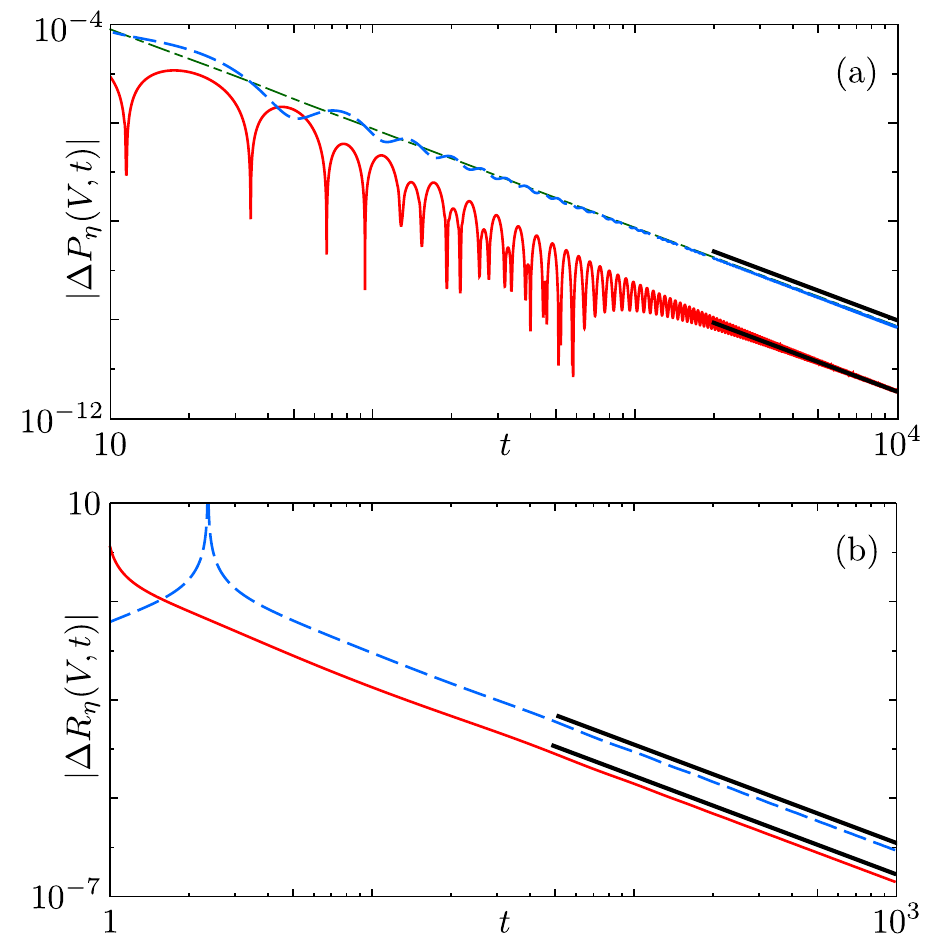}
	\caption{(Color online) Panel (a): Plot of $ |\Delta P_+(V,t)| $ (red, solid), $ |\Delta P_-(V,t)| $ (green, dash-dotted) and $ |\Delta P(V,t)|=|\sum_{\eta}\Delta P_\eta(V,t)| $ (blue, dashed) [units $ \bar{P}_0=|\lambda|^2 q^2_F (\pi^2 K_f)^{-1} $] as function of time $ t $ [units $ (v q_F)^{-1} $] for the quench $ K_i=0.9\rightarrow K_f=0.6 $. Panel (b): Plot of $ |\Delta R_\eta(V,t)|=|R_\eta(V,t)-R_\eta^\infty|$, as a function of time $ t $ [units $ (v q_F)^{-1} $] for the quenches $ K_i=0.9\rightarrow K_f=0.6 $ (blue, dashed) and $ K_i=0.8\rightarrow K_f=0.4 $ (red, solid). Note that $ |\Delta R_\eta(V,t)| $ does not depend on the chirality (see text). In both Panels solid black lines indicate the power law $ \propto t^{-2} $ and $ V=0.1\, v q_F/e $.}
	\label{fig:power}
\end{figure}
Indeed, as shown in Fig.~\ref{fig:power}(a), the non-universal decay mostly affects $\Delta P_+(V,t)$ while $ \Delta  P_-(V,t) $ exhibits an almost perfect decay $ \propto t^{-2} $, which emerges clearly even at short times. Moreover, within the time range we are interested in, one has $|\Delta P_-(V,t)|\gg|\Delta P_+(V,t)|$: The relaxation of the total energy current $ P(V,t)=\sum_{\eta}P_\eta(V,t)$ is thus essentially controlled by the dominant universal decay $ \propto t^{-2} $ of $ P_-(V,t) $. It is worth noting that this leads to an intriguing effect: during the transient, the majority of the \emph{excess} energy current injected from the probe flows in the $ \eta=- $ direction. This is in sharp contrast with the steady-state contribution, which is always dominated by the $\eta=+$ chirality, and with the non-quenched case, where one finds $P_+(V,t) \propto P_-(V,t)$ with $|P_+(V,t)|\geq|P_-(V,t)|$. Moreover, we underline that this effect does not exist for the chiral charge currents, which always satisfy $I_+(V,t)\propto I_-(V,t)$ with $I_+(V,t)>I_-(V,t)$ [see Eq.~\eqref{SM:eq:current} in Appendix~\ref{app:appendix}].
To further exploit this peculiar chirality-dependence of the excess energy current, we inspect the energy fractionalization ratio 
\begin{equation}
R_\eta(V,t)=\frac{P_\eta(V,t)}{\sum_\eta P_\eta(V,t)} 
\end{equation}
in the transient regime~\cite{Karzig:2011,Calzona:2017}. Its relaxation towards the steady-state value $R_\eta^\infty=A_\eta^2/(A_+^2+A_-^2)$ \cite{Karzig:2011,Calzona:2017} is depicted in Fig.~\ref{fig:power}(b), where we show the behavior of the absolute value of $\Delta R_\eta(V,t) = R_\eta(V,t)-R_\eta^\infty$ for two different quenches. The universal decay $\propto t^{-2}$ emerges as clearly as in $\Delta P_-(V,t)$. Note that, since $ R_\eta(V,t)=1-R_{-\eta}(V,t) $, one has $ |\Delta R_\eta(V,t)|=|\Delta R_{-\eta}(V,t)|  $, i.e. $ |\Delta R_\eta(V,t)| $ does not depend on the chirality. Moreover, the fractionalization ratio has the key advantage to be time-independent in the non-quenched case, with $ R_\eta(V,t) = R^\infty_\eta$. Therefore, the presence of a transient in $ R_\eta(V,t) $ is a direct hallmark of the non-equilibrium dynamics of the system induced by the quench. Together with $\Delta P_-(V,t)$, it represents a very promising tool for the investigation of the quench-induced entanglement between counter-propagating chiral fields $\phi_{f,\pm}(x)$ and its relaxation in time. 

Before closing, we note that our results hold even for a non-sudden change of the inter-particle interaction and in the presence of a finite temperature. In particular, since the smallest time scale of the system is set by the time cutoff $\tau_0=a/v $, our discussions remain valid  for protocols faster than $ \tau_0 $. On the other hand, a finite temperature $ T $ induces a characteristic time-scale $ \tau_{\mathrm{th}}=\beta \hbar $, with $ \beta $ the inverse temperature, and thus its effects remain negligible for times $ t\ll \tau_{\mathrm{th}} $. In the specific case of a typical fermionic cold atoms setup one has $\tau_0\sim10^{-7}\, \mathrm{s} $ and $ T \sim 10^{-8} \,\mathrm{K} $~\cite{Inguscio:2007}. The associated time scale is then $ \tau_{\mathrm{th}} \sim 10^{-3} s= 10^{4} \tau_0 $ and is well beyond the time region in which the power-law decay $ \propto t^{-2} $ clearly emerges in the energy current and fractionalization ratio in Fig.~\ref{fig:power}.
	
\section{Conclusions} \label{sec:conclusions}
In summary, the non-equilibrium dynamics of a 1D interacting system after a quantum quench has been discussed. It has been shown that an interaction quench results in an initial entanglement between right- and left-moving density excitations, which is encoded in the time evolution of their cross-correlators. This represents a direct fingerprint of the quantum quench and deeply affects the relaxation towards the steady state of the system.
We have shown this in the specific case of spectral and transport properties of a fermionic 1D system subject to an interaction quench. In particular, we demonstrated that the entanglement dynamics induces a universal long-time decay $ \propto t^{-2} $ in the non-equilibrium spectral function, whose time-evolution is also affected by other non-universal power laws, with quench-dependent exponents. Interestingly,
the universal character clearly emerges by considering charge and energy currents in a transport setup. In particular, fractionalization phenomena, peculiar of 1D interacting systems, can be envisioned to probe the presence of quench-induced entanglement and its relaxation. Among all, the transient dynamics of the energy fractionalization ratio represents a promising tool to observe these universal features. 

We expect our discussions to be independent of the precise form of the quench protocol, of the tunneling Hamiltonian or of the presence of a finite temperature. Moreover, our results can elucidate fundamental aspects of non-equilibrium physics settled by a quantum quench and can be tested with state-of-the-art implementation of cold atomic systems or solid state devices.
\\
	
A. C. and F. M. G. contribute equally to this work.

\appendix
\section{Transport properties and spectral function}\label{app:appendix}
In this Appendix we derive the chiral charge and energy currents and study their long-time behavior (Sec.~\ref{app:transport}). We then connect them with the NESF and inspect the asymptotic behavior of the latter (Sec.~\ref{app:spectral}). 

\subsection{Chiral charge and energy currents} \label{app:transport}
We start from the evaluation of transport properties for the setup introduced in Sec.~\ref{sec:trasport}. We begin by quoting the expression for the average variation $ \langle\delta \mathcal{O}(x,t)\rangle_i $ of a generic Hermitian and particle-number conserving operator $ \mathcal{O}(x,t) $ of the system, induced by the switching on of the tunnel-coupling for $ t>0 $ (see Refs.~\cite{Calzona:2016,Calzona:2017} for further details). We assume the system and the probe to be in thermal equilibrium for $ t<0 $, with $ \rho(0) $ the associated global density matrix at $ t=0 $. Working in the interaction picture with respect to the tunneling Hamiltonian $ H_t(t) $ [see Eq.~\eqref{eq:Ht}], to the lowest order in the tunneling amplitude $ \lambda $ one has
\begin{widetext}
\begin{align}
\langle\delta \mathcal{O}(x,t)\rangle_i &= \text{Tr}\{ \mathcal{O}(x,t)[\rho(t)-\rho(0)] \}\nonumber\\
&=2 \text{Re} \int_0^{t} d\tau_1 \int_0^{\tau_1}d\tau_2\,  \text{Tr}\left\{ \rho(0) H_t^+(\tau_2) \left[\mathcal{O}(x,t), H_t^-(\tau_1) \right] +  \rho(0) H_t^-(\tau_2) \left[ \mathcal{O}(x,t), H_t^+(\tau_1) \right] \right\},
\label{SM:eq:variationO}
\end{align}
\end{widetext}
with $ 	H_t(\tau)=H^+_t(\tau)+H^-_t(\tau)=\vartheta(\tau)\lambda\psi_R^\dagger(x_0)\chi(x_0)+\mathrm{h.c.} $. Note that, since $ \mathcal{O}(x,t) $ is an operator acting only on the system, it commutes with the probe Fermi field $ \chi(x) $. 

In order to evaluate the chiral charge current of Eq.~\eqref{eq:chiralI},
\begin{equation}
 I_\eta(V,t) = e\partial_t \int_{-\infty}^{\infty} \langle\delta n_\eta(x,t)\rangle_i \,dx,
\end{equation} 
we substitute $ \mathcal{O}(x,t)=n_\eta(x,t) $ in Eq.~\eqref{SM:eq:variationO}, with $ n_\eta(x,t) $ given in Eq.~\eqref{eq:chiraldensity}. As a first step we evaluate the correlators of the non-interacting probe, modeled as a non-interacting one-channel LL. Note that its corresponding chemical potential is shifted by the bias energy $ eV $ with respect to the Fermi level of the system. In the zero-temperature case we obtain 
\begin{align}
\langle \chi^\dagger(x_0,\tau_2) \chi(x_0,\tau_1) \rangle&=iG^{>}_p(\tau_1-\tau_2)e^{ieV(\tau_2-\tau_1)},\\	
\langle \chi(x_0,\tau_2) \chi^\dagger(x_0,\tau_1) \rangle&=iG^{>}_p(\tau_1-\tau_2)e^{-ieV(\tau_2-\tau_1)},
\end{align}
where the greater Green function of the local probe is
\begin{equation}
\label{SM:eq:G_p}
G^{>}_p(\tau)=-\frac{i}{2\pi a} \frac{a}{a-i v \tau}.
\end{equation}
Then, we focus on the commutator present in Eq.~\eqref{SM:eq:variationO}, which gives
\begin{align}
\left[n_\eta(x,t),\psi^\dagger_R(x_0,\tau)\right]&=\frac{\sqrt{K_f}A_\eta}{\pi}\left[\frac{a}{a^2+(z_{f,\eta}-\bar{z}_{f,\eta})^2}\right]\nonumber\\ &\times\psi^\dagger_R(x_0,\tau),
\end{align}
with generalized coordinates $ z_{f,\eta}=x-\eta u_f t $ and $ \bar{z}_{f,\eta}=x_0-\eta u_f \tau $. Finally, noting that $ \langle \psi_R^\dagger(x_0,\tau_2) \psi_R(x_0,\tau_1)\rangle_i =\langle\psi_R(x_0,\tau_2)\psi_R^\dagger(x_0,\tau_1)\rangle_i=-iG^{<}_R(\tau_1,\tau_2) $, with $ G^{<}_R(\tau_1,\tau_2) $ given in Eq.~\eqref{eq:Glesseraq}, we obtain 
\begin{align}
I_\eta(V,t)&=I^0_\eta \,\text{Re}\bigg[\int_0^tG^{<}(t,t-\tau)G_p^{>}(\tau)\nonumber\\
&\times i \sin(eV\tau)\, d\tau\bigg], \label{SM:eq:current}
\end{align}
where $ I^0_\eta=e|\lambda|^2(1+\eta K_f) $. Note that in the above equation $ G^{<}(t,t-\tau) $ is always in the regime with  $ t>\tau $. 

We now turn to the long-time behavior of Eq.~\eqref{SM:eq:current}. We observe that the main features of the integrand are located in two well separated regions near the boundary of the integration domain. Indeed, $ G^{>}_p(\tau)$ presents a pole in $ \tau=-i a/v $, $ G^{<}_\infty(\tau)$ has branch points in $\tau=\pm i a/u_f$ and $\mathcal{U}(t,\tau)$ has two further branch points at $\tau=t\pm i a/(2 u_f)$.
Moreover, we note that in the region $ 0 <\tau < t $ both $G^{<}(t,t-\tau)$ and $G^{>}_{p}(\tau)$ are smooth and slowly varying functions. Therefore, due to the presence of the oscillating term, the main contribution to the integral arises from the singular parts of the integrand only. We thus have 
\begin{equation}
\label{eq:sup:Isum}
I_\eta(V,t)\approx I_{\eta,0}(V,t)+I_{\eta,t}(V,t),
\end{equation}
with $ I_{\eta,0}(V,t) $ and $ I_{\eta,t}(V,t) $ the contributions due to regions near $ \tau\sim0 $ and $ \tau\sim t $, respectively.
At first, let us focus on $ I_{\eta,0}(V,t)$. To this end, we can exploit the long-time limit expansion of $ G^{<}(t,t-\tau) $ of Eq.~\eqref{eq:Glesser}, retaining only the lowest order in $ \tau/t $.
Since we are interested in the region near  $ \tau\sim 0 $ and thanks to the oscillating term, we can set $t\to \infty$ in the integration domain, obtaining the leading contribution
\begin{align}
I_{\eta,0}(V,t)&\approx I_\eta^0\text{Re}\left[\int_{0}^{\infty} G^{<}_\infty(\tau)\left(1+\frac{\gamma \tau^2}{2 t^2}\right)G_p^{>}(\tau) e^{ieV\tau}\, d\tau\right]\nonumber\\
&= I_\eta^{\infty}(V)-\frac{\gamma}{2e^2t^2}\frac{d^2}{d V^2} I_\eta^\infty(V),
\label{SM:eq:B0}
\end{align}
with \begin{equation}
\label{eq:sup:Iinfty}
I_{\eta}^{\infty}(V)=I^{0}_\eta\text{Re}\left[\int_{0}^{\infty} G^{<}_\infty(\tau)G_p^{>}(\tau)i \sin(eV\tau)\, d\tau\right]
\end{equation}
the asymptotic value of $ I_{\eta}(V,t) $. Therefore, the universal power-law decay of the system Green function found in Eq.~\eqref{eq:Glesser} results in an analogous behavior in the contribution $ I_{\eta,0}(V,t) $.\\
Let us now discuss the region around $ \tau\sim t $. 
After  the change of variable $y=t-\tau$, the main contribution to $ I_{\eta,t}(V,t) $ is given by the term $ (a^2+4u_f^2y^2)^{-\gamma} $ present in $ \mathcal{U}(t,t-y)$. To get the long-time behavior, we thus expand all other contributions at lowest order in $y/t\ll1$. Again, we can safely extend the integration domain to $t\to \infty$, obtaining
\begin{align}
I_{\eta,t}(V,t)&\approx -\frac{I_\eta^{0}}{2\pi^2 a v}  \left(\frac{a}{u_f}\right)^{\nu_++\nu-}\frac{\cos\left[ \frac{\pi}{2}(\nu_+-\nu_-)\right]}{4^{2\gamma}t^{\nu_++\nu_--2\gamma+1}}\nonumber\\
&\times\int_{0}^{\infty} \frac{\sin\left[eV(t-y)\right]}{[(a/2 u_f)^2+ y^2]^{\gamma}}\,dy.
\end{align}
Therefore, in the long-time limit, $ I_{\eta,t}(V,t) $ is an oscillating function decaying with a power law $ \propto t^{-\nu-1}$, with
\begin{equation}\label{SM:eq:nu}
\nu=\nu_++\nu_--2\gamma=\frac{K_f^4+K_i^2+3K_f^2(1+K_i^2)}{8K_f^2K_i}\geq1~.
\end{equation}
Recalling Eq.~\eqref{eq:sup:Isum}, the tunneling current in the long-time limit thus reads
\begin{equation}
I_\eta(V,t)\approx \bar{I}_0\bigg[ \bar{I}_\eta^{\infty}(V)+\frac{\mathcal{I}_{\eta,2}(V)}{t^2} + \frac{M_\eta^I(V,t)}{t^{\nu+1}}\bigg],
\end{equation}
with $ \bar{I}_0=e|\lambda|^2(2\pi^2a v)^{-1}$, $ \bar{I}_\eta^{\infty}(V)=I^\infty_\eta(V)/\bar{I}_0$ and 
\begin{align}
\mathcal{I}_{\eta,2}(V)&=-\frac{\gamma}{2e^2}\frac{d^2}{dV^2}\bar{I}_\eta^{\infty}(V),\\
M_\eta^{I}(V,t)&= -\frac{1+\eta K_f}{2}  \left(\frac{a}{u_f}\right)^{\nu_++\nu-}\frac{\cos\left[ \frac{\pi}{2}(\nu_+-\nu_-)\right]}{4^{2\gamma}}\nonumber\\
&\times\int_{0}^{\infty} \frac{\sin\left[eV(t-y)\right]}{[a^2/(2 u_f)^2+ y^2]^{\gamma}}\,dy.\label{app:eq:MI}
\end{align}
Since $ \nu\geq 1 $, the long-time behavior of the charge current is controlled by the universal contribution $ \propto t^{-2} $.
Note that, by substituting the complete series expansion of Eq.~\eqref{eq:Glesser} in Eq.~\eqref{SM:eq:B0}, the whole contribution with integer power laws reported in Eq.~\eqref{eq:current} is easily recovered  
\begin{equation}
I_\eta(V,t)=\bar{I}_0\bigg[\bar{I}_\eta^\infty(V)+\sum_{n=2}^{\infty}\frac{\mathcal{I}_{\eta,n}(V)}{t^n}+\frac{\mathcal{M}_\eta^I(V,t)}{t^{\nu+1}}\bigg],
\label{SM:eq:currentfinal}
\end{equation}
where the function $\mathcal{M}^{I}_\eta(V,t)$ takes into account all higher order contributions.

The chiral energy current of Eq.~\eqref{eq:chiralP}, 
\begin{equation}
P_\eta(V,t) = \partial_t \int_{-\infty}^{\infty} \langle\delta \mathcal{H}_\eta(x,t)\rangle \,dx,
\end{equation}
can be obtained from Eq.~\eqref{SM:eq:variationO} by substituting $ \mathcal{O}(x,t)=\mathcal{H}_\eta(x,t) $, with $ \mathcal{H}_\eta(x,t) $ given in Eq.~\eqref{eq:chiralhamiltonian}. In this case the commutator in Eq.~\eqref{SM:eq:variationO} gives
\begin{align}
\left[\mathcal{H}_\eta(x,t),\psi_R^\dagger(x_0,\tau)\right] &= -\frac{\eta u_f A_\eta}{\sqrt{2\pi}}\left[\frac{a}{a^2+(z_{f,\eta}-\bar{z}_{f,\eta})^2}\right] \nonumber\\
&\times\partial_x \left\{ \phi_\eta(z_{f,\eta}), \psi_R^\dagger(x_0,\tau) \right\},
\end{align}
with $ z_{f,\eta}=x-\eta u_f t $ and $ \bar{z}_{f,\eta}=x_0-\eta u_f \tau $. From the above equation it emerges that quantum averages of the form 
\begin{equation}
\langle  \psi_R^\dagger(x_0,\tau_2)  \phi_\eta(z_\eta)  \psi_R (x_0,\tau_1) \rangle_i \label{SM:3pointaverage}
\end{equation}
have to be computed. This can be done by using the bosonization identity of Eq.~\eqref{eq:bosonization}, the relation
\begin{equation}
\phi_\eta(z_\eta) = -i \partial_y e^{i y  \phi_\eta(z)}\Big|_{y = 0}
\end{equation}
and the well-known properties of gaussian averages valid for bosonic fields~\cite{Voit:1995,vonDelft:1998,Giamarchi:2004}. We obtain
\begin{align}
P_\eta(V,t)\!&=\!P^0\text{Re}\bigg\{\!\int_{0}^{t}\!\left[\frac{1}{i u_f}\!\left(\partial_{\bar{t}}-\eta u_f\partial_\xi\right)\!G_R^{<}(\xi;\bar{t},t-\tau)\right]_{\substack{\bar{t}=t \\ \xi=0}}\nonumber\\
&\times G^{>}_{p}(\tau)\cos(eV \tau) \,d\tau\bigg\},
\label{SM:eq:energy}
\end{align}
with $ P^0=2 |\lambda|^2 u_f$ and $ G_R^{<}(\xi;\bar{t},t-\tau)=i\langle\psi_R^{\dagger}(0,t-\tau)\psi_R(\xi,\bar{t})\rangle $ the non-local Green function. Again, the properties of $ P_\eta(V,t) $ can be expressed in terms of two-point correlators. Indeed, the term in square brackets evaluates to
\begin{multline}
\left[\frac{1}{i u_f}\left(\partial_{\bar{t}}-\eta u_f\partial_\xi\right)G_R^{<}(\xi;\bar{t},t-\tau)\right]_{\substack{\bar{t}=t \\ \xi=0}}\\
=G^{<}(t,t-\tau)\mathcal{F}_{\eta}(t,\tau),
\end{multline}
where
\begin{equation}
\mathcal{F}_\eta(t,\tau)=A^2_\eta\mathcal{F}_1(\tau)+\gamma\mathcal{F}_2(t,\tau)
\end{equation}
and 
\begin{align}
\mathcal{F}_1(\tau)&=\frac{\theta_+^2}{a-i u_f \tau}-\frac{\theta_-^2}{a+i u_f \tau},\label{SM:eq:F1}\\
\mathcal{F}_2(t,\tau)&=\frac{2iu_f(2t-\tau)}{a^2+u^2_f(2t-\tau)^2}-\frac{4iu_ft}{a^2+4u^2_ft^2}.
\label{SM:eq:F2}
\end{align}
The long-time behavior of Eq.~\eqref{SM:eq:energy} can be obtained following the same lines  illustrated in the case of the charge current. We therefore arrive at
\begin{equation}
P_\eta(V,t)\!=\!\bar{P}_0\bigg[\bar{P}^\infty_\eta(V)\!+\!\sum_{n=2}^{\infty}\frac{\mathcal{P}_{\eta,n}(V)}{t^n}\!+\!\frac{\mathcal{M}^P_\eta(V,t)}{t^{\nu+2}}\bigg],
\end{equation}
with $ \bar{P}_0=|\lambda|^2 (\pi^2 a^2 K_f )^{-1} $ and 
\begin{widetext}
\begin{align}
\bar{P}^{\infty}_\eta&=2\pi^2 a^2 v A^2_\eta\text{Re}\left[ \int_{0}^{\infty} G^{<}_\infty(\tau)\mathcal{F}_1(\tau)G_p^{>}(\tau)\cos(eV\tau)\,d\tau\right],\\
\mathcal{P}_{\eta,2}(V)&=-\gamma\left\{\frac{1}{2}\frac{d^2}{e^2dV^2}\bar{P}^{\infty}_\eta(V)- 2\pi^2 a^2 K_f \text{Re}\left[\int_{0}^{\infty}G^{<}_\infty(\tau)G_p^{>}(\tau)\cos(eV\tau)i\tau \,d\tau\right]\right\},\\
\mathcal{M}_\eta^P(V,t)&\approx- [A_\eta^2(\theta^2_++\theta^2_-)+\gamma]\left(\frac{a}{u_f}\right)^{\nu_++\nu_-+1} \frac{\cos\left[\frac{\pi}{2}(\nu_+-\nu_-)\right]}{4^{2\gamma}} \int_{0}^{\infty}\frac{\cos[eV(t-y)]}{\left[a^2/(2u_f)^2+y^2\right]^\gamma}\, dy, \label{app:eq:MP}
\end{align}
\end{widetext}
where in the last line we have retained only the leading order in the long-time limit expansion.  
\subsection{Spectral function}\label{app:spectral}
We start this Section by pointing out the explicit connection between the charge and energy currents with the system lesser NESF of Eq.~\eqref{eq:spectral}, 
\begin{equation}
\label{eq:sup:A<}
A^{<}(\omega,t)\equiv \frac{1}{2\pi}\int_{-\infty}^{\infty}e^{i\omega \tau}(-i)G^{<}(t,t-\tau)\,d\tau.
\end{equation} 
One can rewrite Eqns.~\eqref{SM:eq:current} and~\eqref{SM:eq:energy} as
\begin{align}
I_\eta(V,t)&=I_\eta^0\mathrm{Re}\left[\int_{-\infty}^{\infty} A^{<}(\omega,t)\mathcal{B}_{V}^{I}(\omega,t)\,d\omega\right],\\
P_{\eta}(V,t)&=P^0\text{Re}\left[\int_{-\infty}^{\infty}A^{<}(\omega,t)\mathcal{B}^{P}_{V,\eta}(\omega,t)\,d\omega\right],
\end{align} 
with the functions
\begin{align}
\mathcal{B}^{I}_{V}(\omega,t)&=\sum_{\eta=\pm}\eta\int_{-\infty}^{\infty} A_{p}^{>}(\Omega) \mathcal{S}_{\eta V}(\omega,\Omega,t)\,d\Omega,\\
\mathcal{B}_{V,\eta}^{P}(\omega,t) &= \sum_{\ell=\pm}\iint_{-\infty}^{\infty} A^{>}_p(\Omega) \widetilde{\mathcal{F}}_\eta(\omega',t) \nonumber\\
&\times\mathcal{S}_{\ell V}(\omega+\omega',\Omega,t)\,d\omega' d\Omega.
\end{align}
Here, we have introduced the greater spectral function of the probe $ A^{>}_{p}(\Omega)=(2\pi)^{-1}\int_{-\infty}^{\infty}e^{i\Omega\tau}iG^{>}_{p}(\tau) \, d\tau $ as well as functions
\begin{align}
\mathcal{S}_{\eta V}(\omega,\Omega,t)&=\frac{i\left[1-e^{i(\eta eV-\omega-\Omega)t}\right]}{\nu eV-\omega-\Omega},\\
\widetilde{\mathcal{F}}_\eta(\omega',t)&=\frac{1}{2\pi}\int_{-\infty}^{\infty}\mathcal{F}_\eta(t,t-\tau)e^{i\omega' \tau}\, d\tau.
\end{align}

We now discuss the asymptotic behavior of the NESF of Eq.~\eqref{eq:sup:A<} and derive the expansion of Eq.~\eqref{eq:Alesser}. In this case, in contrast with the previous Section, one should also consider the region with $t<\tau$. Following the same arguments of Sec.~\ref{sec:spectral}, one can demonstrate that in this regime
\begin{equation}
G^{<}(t,t-\tau)=\mathscr{G}(t,\tau)\mathscr{U}(t),\label{SM:eq:Glesserbq}
\end{equation}
with
\begin{align}
\mathscr{G}(t,\tau)&=\frac{i}{\pi a}\prod_{\ell=\pm}\left\{ \frac{a}{a+i[(u_i+\ell u_f)t-u_i\tau]} \right\}^{\nu_\ell-2\gamma},\\
 \mathscr{U}(t)&=\left(\frac{a^2}{a^2+4u_f^2t^2}\right)^{\gamma}.
\end{align}
In order to obtain the long-time behavior of $ A^{<}(\omega,t) $ we exploit the same method used for charge and energy currents in Section~\ref{app:transport}. Note that the integrand function $ G^{<}(t,t-\tau) $ present in Eq.~\eqref{eq:sup:A<} has singular points in $ \tau=\pm i a/u_f $, $ \tau=t\pm ia/(2u_f) $ and $ \tau=(1+u_f/u_i)t-i a/u_i $.
We thus have
\begin{equation}
\label{app:eq:AlesserLT}
A^{<}(\omega,t)\approx A^{<}_0(\omega,t)+A^{<}_t(\omega,t)+A^{<}_{(1+u_f/u_i)t}(\omega,t).
\end{equation}
\\
In particular, using the long-time expansion of $ G^{<}(t,t-\tau) $ in Eq.~\eqref{eq:Glesser}, we obtain that the region near $ \tau\sim0 $ gives, as in previous cases, the asymptotic contribution and the universal power-law decaying one. On the other hand, regions near $ \tau\sim t $ and $ \tau\sim(1+u_f/u_i)t $ result in two slightly different non-integer power-law decays. Following the same steps leading to Eq.~\eqref{SM:eq:currentfinal} we thus obtain
\begin{equation}
A^{<}(\omega,t)=\bar{A}_0\bigg[\bar{A}^{<}_\infty(\omega)+\sum_{n=2}^{\infty}\frac{\mathcal{A}_n(\omega)}{t^n}+\frac{\mathcal{M}^{A}(\omega,t)}{t^{\nu}}\bigg],
\end{equation}
with $ \bar{A}_0=(2\pi^2 v)^{-1} $ and 
\begin{align}
\bar{A}^{<}_{\infty}(\omega)&=\pi v \int_{-\infty}^{\infty}e^{i \omega \tau}(-i) G^{<}_\infty(\tau)\,d\tau,\\
\mathcal{A}_2(\omega)&=-\frac{\gamma}{2}\frac{d^2}{d\omega^2}\bar{A}^{<}_{\infty}(\omega),\\
\label{eq:sup:MA}
\mathcal{M}^A(\omega,t)&\approx\frac{v}{a} \frac{e^{i\frac{\pi}{2}(\nu_--\nu_+)}}{4^{2\gamma}}\left(\frac{a}{u_f}\right)^{\nu_++\nu_-}e^{i\omega t}\nonumber\\
&\left\{\int_{0}^{\infty}e^{-i\omega y}\frac{1}{[a^2/(2u_f)^2+y^2]^{\gamma}}\, dy\right.\nonumber\\
&\left.+\frac{2^{2\gamma}}{i\omega}\left(\frac{u_f}{a}\right)^{2\gamma} \right\}.
\end{align}
We point out that in Eq.~\eqref{eq:sup:MA} we have retained only the leading order in the time expansion; the first term in curly brackets stems from the regime $ \tau<t $, while the second one is due to the regime $ \tau>t $. Finally, we note that in Eq.~\eqref{app:eq:AlesserLT} all the three contributions inside the square brackets are dimensionless.


\begin{thebibliography}{63}
\bibitem{footnote:g2g4} For the sake of simplicity we focus on the Galilean invariant case only, with $ g_{\mu,2}=g_{\mu,4}\equiv g_{\mu} $. The more general case with $ g_{\mu,2}\neq g_{\mu,4} $ can be obtained straightforwardly without any significant change.
\bibitem{footnote:direct} In the following we will consider only quenches which increase the strength of inter-particle interactions, i. e. with $ K_i>K_f $.
\bibitem{footnote:autocorrelators} Note that the relation $ D_{\alpha,\alpha}(\xi;t,\tau)\equiv D_{\alpha,\alpha}(\xi;\tau) $ holds $ \forall \tau<t $ and not only in the long-time limit.
\bibitem{footnote:nonlocal} Note that, due to symmetric combination of cross-correlators, the leading time-dependent contribution to $ G^{<}(t,t-\tau) $ would be $ \propto t^{-2} $ also in the non-local case.
\bibitem{footnote:cutoff} Here and in the following we make the identification $ a=q_F^{-1} $.
\end{thebibliography}
\end{document}